\documentclass[prd,twocolumn,preprintnumbers,superscriptaddress,nofootinbib]{revtex4}
\usepackage[a4paper, hdivide={1.91cm,,1.165cm}, vdivide={1.83cm,,2.6cm}]{geometry}
\usepackage{amstext,amssymb}
\usepackage{amsmath}
\usepackage{graphicx}
\usepackage[hyperfootnotes=false]{hyperref}
\usepackage{xspace}
\usepackage{color}
\usepackage{units}
\usepackage{slashed} 
\usepackage{braket}

\usepackage{wrapfig}
\definecolor{light-gray}{gray}{0.95}
\usepackage{tcolorbox}

\definecolor{bostonuniversityred}{rgb}{0.8, 0.0, 0.0}

\makeindex

\usepackage{scalerel}
\usepackage{soul}

\usepackage{hyperref}
\hypersetup{ 
	setpagesize=false,
	bookmarksnumbered=true,
	colorlinks=true,
	linkcolor=blue,
	citecolor=red,
	hypertexnames=true
}

\begin{document}

\title{\hspace*{-1.0cm} Zoom in muon survival probability with sterile neutrino for CP and T-violation}


\author{Kiran \surname{Sharma}}
 \email{kirans@iitbhilai.ac.in}
\affiliation{Department of Physics, Indian Institute of Technology Bhilai, India}

\author{Sudhanwa \surname{Patra}}
 \email{sudhanwa@iitbhilai.ac.in}
\affiliation{Department of Physics, Indian Institute of Technology Bhilai, India}

\begin{abstract}
We present the approximated analytic expressions for the muon survival probability in a $3+1$ mixing scenario in the presence of matter effect using the S-matrix formalism. We find that all the individual terms contributing to the muon survival probability can significantly reduce to just three contributions. The leading order contribution comes from the three flavor muon survival probability followed by the two sub-leading contributions arising from active-sterile mixing. Furthermore, to more simplify the results we adopt the well known series expansion relations about mass-hierarchy parameter $\alpha = \Delta m^2_{21} / \Delta m^2_{31}$ and the mixing angle $\sin \theta_{13}$ in the vanishing limit of $\alpha^2$. We discuss the relevance of muon survival probability to probe the CP and T-violation studies coming from the new physics. We also compare the analytic relation between vacuum and matter contributions to the muon survival probability at the leading order. Finally, we comment on the probability behavior at the various long baselines relevant to understand the atmospheric-neutrino sector and to resolve the existing mass-hierarchy problem.
\vspace*{-1.0cm}
\end{abstract}


\maketitle
\noindent
\section{Introduction}
\label{sec:intro}

The famous solar-neutrino puzzle~\cite{Bethe:1986ej,SNO:2001kpb} and the atmosphere neutrino problem~\cite{Super-Kamiokande:1998kpq} leads to the establishment of the phenomenon of neutrino oscillations~\cite{Bilenky:1998dt, Giunti:2004pd}. These oscillations occur among the three active neutrino flavor states $\nu_e$, $\nu_\mu$ and $\nu_\tau$. The oscillation parameters in the $3$ flavor mixing scenario include the solar mass-square difference $\Delta m^2_{21}$, the atmospheric mass-square difference $\Delta m^2_{31}$, three mixing angles $\theta_{12}$, $\theta_{13}$ and $\theta_{23}$ and one leptonic CP-phase $\delta_{13}$. The neutrino oscillations are feasible only if at least one of three neutrino mass eigen state is non-degenerate and the mixing angles are non-zero. Thus neutrinos have a non-zero mass contrary to the standard model prediction. With the discovery of neutrino oscillations, revolutionary research begins looking for the beyond standard model physics. In the current precision era, the main goals of current and proposed experiments are to solve the unknowns in neutrino oscillations which include the mass hierarchy problem, the octant issue, and determination of the CP-violation phase. Moreover, with time certain anomalies~\cite{LSND:2001aii,MiniBooNE:2018esg,Mention:2011rk,GALLEX:1997lja} have emerged which hints toward the existence of a fourth sterile neutrino state. Including the presence of such sterile neutrino along with the three active neutrinos increases the number of oscillation parameters, we have the sterile-active neutrino mixing angles ($\theta_{14}$ ,$\theta_{24}$, $\theta_{34}$), the new mass-squared differences ($\Delta m^2_{41}$, $\Delta m^2_{42}$, $\Delta m^2_{43}$) and the new CP-violating phases ($\delta_{14}$, $\delta_{24}$ and $\delta_{34}$). The presence of sterile neutrino may answer the smallness of neutrino mass. 

The probability level analysis in the presence of sterile neutrinos has already been carried out in the literature~\cite{Giunti:2009xz,Abazajian:2012ys,Palazzo:2013me,Kopp:2013vaa, Gariazzo:2015rra,Giunti:2015wnd, Giunti:2015jba,Choubey:2017cba,Agarwalla:2018nlx,Giunti:2019aiy,Boser:2019rta,T2K:2019efw}. In the reference~\cite{Klop:2014ima}, authors have given transition probability $P^{4\nu}_{\mu e}$ in the $3+1$ scheme by carrying out the S-matrix analysis. In our recent work~\cite{Sharma:2022qeo}, we have also given a slightly different formalism which simplifies the transition probability $P^{4\nu}_{\mu e}$ and survival probability $P^{4\nu}_{e e}$ using One-scale mass dominance(OMSD)~\cite{Choubey:2003yp}. In the present work, we provide the analytic expression for muon survival probability in the presence of matter potential~\cite{Mikheyev:1985zog,Wolfenstein:1977ue,Smirnov:2004zv}. Using the S-matrix formalism, we decompose the $4 \times 4$ Hamiltonian to the effective $3 \times 3$ Hamiltonian. We perform the series expansion~\cite{Akhmedov:2004ny} about small parameters $s_{13} \approx \mathcal{O}(\epsilon)$ and $\alpha \approx \mathcal{O}(\epsilon^2)$  such that the final probability expression for muon appearance probability is of  order $\mathcal{O}(\epsilon^3)$. We discuss the term-by-term analysis of leading and sub-leading contributions to the final muon survival probability. The analytic calculations involved in $P^{4\nu}_{\mu \mu}$ greatly simplify
 to just $3$ significant contributions. We emphasis on the advantage of muon survival probability for the study of CP violation~\cite{Arafune:1997hd,Bilenky:1997dd,Dighe:2008bu,Gandhi:2015xza,Petcov:2018zka} and T-violation~\cite{Cabibbo:1977nk,Kuo:1987km, Parke:2000hu, Akhmedov:2001kd,Xing:2013uxa,Rout:2017udo, Petcov:2018zka,Schwetz:2021thj,Schwetz:2021cuj} in context to $T2K$ experiment~\cite{T2K:2001wmr,T2K:2011qtm,T2K:2021xwb} as a case study. We also look at the probability level analysis for the long baseline, which can shed light on the mass-hierarchy sensitivity.
 
The paper is organized as follows: In the next section we give the analytic expression for muon survival probability in vacuum followed by its treatment in the presence of matter in section $\rm III$. In section $\rm IV$, we provide a formalism which gives the terms contributing to total muon survival probability. In the next leading section, we carry out the well known $\alpha - s13$ approximation and perform the numerical analysis for different baselines. The conclusion is marked in section $\rm VI$, followed by appendices.
 
\noindent
\section{Muon survival probability expression in Vaccum}
\label{sec:Pmuu-vac}
In $3+1$ framework, we have three flavor states for active neutrinos $\nu_e, \nu_\mu, \nu_\tau$ and additional sterile neutrino flavor state $\nu_s$. Also there exists three mass eigenstates $\nu_1, \nu_2, \nu_3$ corresponding to the light neutrinos with masses $m_1, m_2, m_3$ plus one more extra mass eigenstate $\nu_4$ with mass $m_4$. The time evolution of these states in $3+1$ framework is defined as,
\begin{equation}
	i\frac{\partial}{d t}
	\begin{pmatrix}
	 |\nu_e \rangle \\
	 |\nu_\mu \rangle \\
	 |\nu_\tau \rangle \\
	 |\nu_s \rangle 
	\end{pmatrix} = U
\left[ \begin{pmatrix} 0 & 0 & 0 & 0  \\
                          0 & \frac{\Delta m^2_{21}}{2E} & 0 & 0 \\
                          0 & 0 &  \frac{\Delta m^2_{31}}{2E} & 0 \\
                          0 & 0 & 0 &  \frac{\Delta m^2_{41}}{2E}  
       \end{pmatrix}
\right]
U^\dagger \,
\begin{pmatrix}
	 |\nu_e \rangle \\
	 |\nu_\mu \rangle \\
	 |\nu_\tau \rangle \\
	 |\nu_s \rangle 
	\end{pmatrix} \nonumber 
\end{equation}
where $\Delta m^2_{21}$ is the difference between square of masses between different neutrino mass eigenstates and $E$ is the typical energy of the neutrinos of interest. 
The mixing matrix for $3+1$ framework is parametrized in terms mixing angles $\theta_{i4}$ (with i=1,2,3), $\theta_{12}, \theta_{13}$, $\theta_{23}$ and CP-phases. It is important to know the correct parametrization of the neutrino mixing matrix for $3+1$ framework. Although the choice of parametrization can differ depending upon the placement of the CP-phase but the physics will remain same independent of choice of parametrization. The standard parametrization considered for the present work is given by 
\begin{eqnarray}
U &=& R\big(\theta_{34}, \delta_{34} \big) \, R\big(\theta_{24}, 0 \big) \, R\big(\theta_{14}, \delta_{14} \big) \nonumber \\
 &&\hspace*{1cm} \times R\big(\theta_{23}, 0 \big) R\big(\theta_{13}, \delta_{13}\big) R\big(\theta_{12}, 0 \big) 
\end{eqnarray}
Here, $\theta_{14}$, $\theta_{24}$ and $\theta_{34}$ are the mixing angles between sterile and active neutrinos while $\theta_{23}$, $\theta_{13}$ and $\theta_{12}$ are known mixing angles of three flavor neutrinos. On top of standard CP-phase $\delta_{13}$ we have two more CP-phases $\delta_{14}$ and $\delta_{34}$. For completeness,the element-wise description of the unitary mixing matrix is mentioned in the appendix~\ref{app:U44}

The muon survival probability in vaccum using the general expression for the oscillation probability~\cite{Giunti:2007ry} for $N$ flavor neutrinos is found to be, 
\begin{align} \label{N_gen}
 P_{\mu \mu} &= 1 - 4 \sum_{i<j} \text{Re} \big( U_{\mu i} U_{\mu j} U_{\mu j}^* U_{\mu i}^* \big) \sin^2 \bigg(\Delta_{ij}L/4E\bigg) \nonumber \\
                  & + 2 \sum_{i<j}  \text{Im} \big( U_{\mu i} U_{\mu j} U_{\mu j}^* U_{\mu i}^* \big) \sin\bigg(2 \Delta_{ij}L/4E\bigg),
 \end{align}
 where $\Delta_{ij}=m_i^2 - m_j^2$ is the mass-square difference between any of the two mass eigenstates of the neutrinos. For $3+1$ scenario and using the unitarity condition, we get 
 $$\text{Im} \big( U_{\mu i} U_{\mu j} U_{\mu j}^* U_{\mu i}^* \big) 
  =\text{Im} \big( \big|U_{\mu i}\big|^2 \big|U_{\mu j} \big|^2 \big) =0\,. $$
Also, with the presence of eV scale sterile neutrino, the mass-square difference terms are related as
$$\Delta_{41}\simeq \Delta_{42} \simeq \Delta_{43}\,,\,
\mbox{and}\,\Delta_{32}\simeq \Delta_{31}\,. $$
It is to be noted that the terms involving solar mass square difference $\Delta m^2_{21} =\Delta_{21}$ contributes negligible to the frequency part of the survival probability in vaccum and hence, can be dropped from the probability expression. With these simplifications, the muon survival probability in $3+1$ scenario in vaccum reduces to
\begin{align} \label{N_gen}
 P^{(3+1)}_{\mu \mu} &= 1 - 4\, |U_{\mu 3}|^2\,  \big(1+ |U_{\mu 3}|^2 - |U_{\mu 4}|^2 \big)\, 
 \sin^2\bigg(\frac{\Delta_{31}\,L}{4\,E} \bigg) \nonumber \\
 &- 4\, |U_{\mu 4}|^2\,  \big(1 - |U_{\mu 4}|^2 \big)\,
 \sin^2\bigg(\frac{\Delta_{41}\,L}{4\,E} \bigg) 
 \end{align} 
By looking at the muon survival probability expression, it is interesting to note that the long baseline experiments like $T2K$, $NOvA$ etc, can shed light on sterile neutrino parameters like $\theta_{24}$ and $\Delta m^2_{41}$. Thus, the study of survival probability is equally as crucial as transition probability. However, the derived results are valid for vacuum but can not hold true if we include matter effects. So we focus, in the present work, on the analytic derivation of muon survival probability and its phenomenological aspect within the $3+1$ neutrino oscillation scenario in the presence of matter effects.  
 
\noindent
\section{General Treatment of $3+1$ neutrino oscillation in Matter}
\label{sec:Pmuu-vac} 
The neutrino oscillation probabilities can be modified significantly in presence of matter due to interactions between neutrinos and the medium particles. For example, the electron neutrino $\nu_{e}$ and muon neutrino $\nu_\mu$ can scatter with the medium particles due to presence of electrons $e^-$, protons $p$ and neutrons $n$ via charge and neutral current effects. Thus,we must include matter effects on the oscillation probabilities for neutrinos propagating via dense medium such as Earth. The effect is known as Mikheyev-Smirnov-Wolfenstein (MSW) effect~\cite{Mikheyev:1985zog,Wolfenstein:1977ue,Smirnov:2004zv} and resonance enhancement of oscillations can be important for long baseline experiments like T2K, NOvA, and DUNE as well as for atmospheric neutrino detectors like Super-K, INO, and IceCube. 
The present discussion will carry forward basic MSW effect in presence of extra sterile neutrino while adopting various simplifications considered in refs~\cite{Akhmedov:2004ny, Klop:2014ima, Asano:2011nj,Sharma:2022qeo,Choubey:2017cba, Agarwalla:2018nlx}.
\noindent
The neutrino oscillation in presence of matter due to the coherent forward scattering of neutrinos with the medium particles is characterized by potential term $V_{\rm CC}$ as
$$V_{\rm CC} = \sqrt{2}\,G_F\,N_e\,.$$
where $G_F$ denotes the Fermi coupling constant, and $N_e$ is the electron number density inside the matter. The general Hamiltonian for the $3+1$ scenario in the presence of matter is of the following form,
\begin{eqnarray}
H_{4\nu} \;&=&\; 
\underbrace{
U
\left[ \begin{array}{cccc} 0 & 0 & 0 & 0  \\
                          0 & \Delta m^2_{21}/2E & 0 & 0 \\
                          0 & 0 & \Delta m^2_{31}/2E & 0 \\
                          0 & 0 & 0 & \Delta m^2_{41}/2E  
       \end{array}
\right]
U^\dagger}_{\displaystyle =H_{\rm vac}} \nonumber \\
&+&\underbrace{
\left[ \begin{array}{cccc} V_{CC} & 0 & 0 & 0 \\
                          0 & 0 & 0 & 0 \\
                          0 & 0 & 0 & 0 \\
                          0 & 0 & 0 & -V_{NC}
       \end{array}
\right]}_{\displaystyle =H_{\rm mat}} \;, \nonumber \\
&=& U K U^\dagger + V \, .
\label{H4nu}
\end{eqnarray}
where U is the unitary mixing matrix involving three active neutrinos and a sterile neutrino with the form
\begin{eqnarray}
U &=& R\big(\theta_{34}, \delta_{34} \big) \, R\big(\theta_{24}, 0 \big) \, R\big(\theta_{14}, \delta_{14} \big) \nonumber \\
&&\hspace*{2cm} R\big(\theta_{23}, 0 \big) R\big(\theta_{13}, \delta_{13}\big) R\big(\theta_{12}, 0 \big) \nonumber \\
 &\equiv& R\big(\theta_{34}, \delta_{34} \big) \, R\big(\theta_{24}, 0\big) \, R\big(\theta_{14}, \delta_{14} \big) U_{3\nu} 
\end{eqnarray}

It is to be noted that $W$~boson mediated CC interaction potential $V_{CC}$ has played a crucial role in oscillation probability. In contrary, $Z$~boson mediated NC potential $V_{\rm NC}$ is same for all flavors and hence, is phased out from the probabilities. Since the sterile neutrino has no interaction with the medium  particles, the $V_{\rm NC}$ can not be phased out and will play an important role when we consider the $3+1$ scenario of neutrino oscillation. The diagonal matrix 
$K = \mbox{diag}\left(0,k_{21}, k_{31}, k_{41} \right)$ with $k_{i1} = \Delta m^2_{i1}/2E$ is for vacuum part of the Hamiltonian in mass basis in a reduced form.

Let us introduce a change of neutrino flavor basis in $3+1$ framework as
$$\overline{\nu} = \overline{U}^\dagger \nu $$
where the modified mixing matrix in the new flavor basis is read as
\begin{itemize}
 \item The form of $\overline{U}$ is given by 
$$\overline{U} = R\big(\theta_{34}, \delta_{34} \big) \, R\big(\theta_{24}, 0 \big) \, R\big(\theta_{14}, \delta_{14} \big)$$
 \item The $4 \times 4$ Hamiltonian can be decomposed as $3 \times 3$ effective Hamiltonian~\cite{Klop:2014ima}
 \begin{eqnarray}
 \overline{H}_{4\nu} &=& \overline{H}^{\rm kin} + \overline{H}^{\rm dyn} = U_{3\nu} K U^\dagger_{3\nu} + \overline{U}^\dagger V \overline{U} \nonumber
\end{eqnarray}
with $U_{3\nu}= R\big(\theta_{23}, 0 \big) R\big(\theta_{13}, \delta_{13}\big) R\big(\theta_{12}, 0 \big)$.
\end{itemize}
It is to be noted that the first term describes the kinetic contribution while relevant for neutrino oscillation in vacuum while the second term is the dynamic contribution accommodating matter effects. The fact that the $k_{41}$ is very large as compared to other mass square difference terms, the (4,4) entries of $\overline{H}$ is much larger than all other elements and hence, can be decoupled from the $3+1$ framework leading to an effective three flavor analysis. So the fourth eigenstate evolves independently of the other flavor states. Thus, the $4\times4$ complete Hamiltonian in $3+1$ framework can be reduced to an effective projected Hamiltonian in  
three flavor scenario. 
\begin{itemize}
 \item The dynamical contributions after some simplifications is described as,
\begin{eqnarray}
\overline{H}^{\rm dyn} &=& \overline{U}^\dagger V \overline{U} \nonumber \\
& \simeq &V_C \begin{pmatrix}
1-(1-r) s^2_{14}  & r \tilde{s}_{14} s_{24} & r \tilde{s}_{14} \tilde{s}^*_{34} \\
\dagger  & r s^2_{24}  & r s_{24} \tilde{s}^*_{34}  \\
\dagger  & \dagger  & r s^2_{34}
                           \end{pmatrix}
\end{eqnarray}
The factor $r$ is the ratio of neutral current and charged current matter potentials with a negative sign. For the Earth matter, it has a typical value of $0.5$
 \item The S-matrix evolution operator in terms projected Hamiltonian in three flavor is as follows
\begin{eqnarray}
\overline{S} \;&=&\; \, \pmb{e}^{- i\, \overline{H}\,L} \approx 
\left[ \begin{array}{cc} \overline{S}_{3\nu} & {\bf 0}_{3 \times 1} \\
                          {\bf 0}_{1 \times 3} &  \exp\!\left(- i\, k_{41}\,L\right)
       \end{array}
\right]
\label{HinMatter}
\end{eqnarray}
 \item One can revert back to the original flavor  basis by unitary transformations $S = \overline{U}\, \overline{S}\, \overline{U}^\dagger$.\item The probability expression can be readily obtain from the evolution matrix as 
 \begin{equation}
 P^{4\nu}_{\alpha \beta} \equiv P^{4\nu}_{\alpha \beta} \big(\nu_\alpha \to \nu_\beta;L \big)= \big|S_{\beta \alpha} \big|^2
 \end{equation}
\end{itemize}
\noindent
\section{Muon survival probability in $3+1$ scenario}
\noindent
Using the general treatment mentioned in section~\ref{sec:Pmuu-vac}, we evaluate the muon survival probability, with $\overline{U}_{e2} = \overline{U}_{e3} = \overline{U}_{\mu 3} = 0$ (exactly as per parametrized), the relevant component $S_{\mu \mu}$ is given by
 \begin{eqnarray}
  S_{\mu\mu} &=&  \bigg[\big|\overline{U}_{\mu 1}\big|^2 \overline{S}_{ee}+
    \overline{U}^*_{\mu 1} \overline{S}_{\mu e} \overline{U}_{\mu 2}
     +\overline{U}_{\mu 1} \overline{S}_{ e \mu} \overline{U}^*_{\mu 2}
     + \big|\overline{U}_{\mu 2} \big|^2 \overline{S}_{\mu \mu}
    \bigg] \nonumber \\
    &&+ \overline{U}_{\mu 4} \overline{U}^*_{\mu 4} \overline{S}_{ss}
    \nonumber \\
    &=& \mathcal{A} + \mathcal{B}
 \end{eqnarray}
 \newline
 \noindent
and the complex conjugate part $S^*_{\mu\mu}$ is described as
 \begin{eqnarray}
  S^*_{\mu\mu} &=&  \bigg[\big|\overline{U}_{\mu 1}\big|^2 \overline{S}_{ee}+
    \overline{U}^*_{\mu 1} \overline{S}_{\mu e} \overline{U}_{\mu 2}
     +\overline{U}_{\mu 1} \overline{S}_{ e \mu} \overline{U}^*_{\mu 2}
     + \big|\overline{U}_{\mu 2} \big|^2 \overline{S}_{\mu \mu}
    \bigg]^* \nonumber \\
    &&+ \overline{U}^*_{\mu 4} \overline{U}_{\mu 4} \overline{S}^*_{ss}
        \nonumber \\
    &=& \mathcal{A}^* + \mathcal{B}^*
 \end{eqnarray}
\noindent
The disappearance probability is expressed in terms of evolution matrix and other mixing matrices as follows
 \begin{eqnarray}
P^{4\nu}_{\mu \mu} &=& S_{\mu \mu} \cdot S^*_{\mu \mu} \nonumber \\
    &=& \bigg(\mathcal{A} + \mathcal{B} \bigg) \bigg(\mathcal{A}^* + \mathcal{B}^* \bigg)  \nonumber \\
    &=& \mathcal{A} \mathcal{A}^* + \mathcal{A} \mathcal{B}^* 
        + \mathcal{B} \mathcal{A}^* + \mathcal{B} \mathcal{B}^*
 \end{eqnarray}
 \noindent
 Averaging out terms containing $\big|\overline{S}_{ss} = e^{-i\, k_{41} L}\big|^2$ gives a factor of 1/2 while terms containing only $\overline{S}_{ss}$ can be averaged out completely from the general expression giving vanishing effects. The terms containing only $\mathcal{B}$ or $\mathcal{B}$ (carries the sterile factor $\overline{S}_{ss}$) are vanishing. The other factor $\mathcal{B} \mathcal{B}^*$ results
 \begin{eqnarray}
  \mathcal{B} \mathcal{B}^* = \big|\overline{U}_{\mu 4} \big|^4 \cdot \frac{1}{2} \nonumber \\
  &\approx& \frac{1}{2} \cos^4\theta_{14} \sin^4\theta_{24} \nonumber \\
  &\simeq& \frac{1}{2} \sin^4\theta_{24} \quad \mbox{of order} \quad  \mathcal{O}(\epsilon^4) 
   \end{eqnarray}
 This term has negligible contribution as we keep only terms upto order $\mathcal{O}(\epsilon^3)$ in the probability expression. 
\noindent 
 The only relevant terms after averaging out terms involving $\Delta m^2_{41}$ are presented below
\begin{eqnarray}
P^{4\nu}_{\mu \mu}
    &=& \mathcal{A} \mathcal{A}^* = \sum^{16}_{k=1} T_k 
\end{eqnarray}   
while the individual terms are expressed as,
 \begin{widetext}
 \begin{eqnarray}
&&\hspace*{-0.5cm} T_{1} =  
 \big|\overline{U}_{\mu 1} \big|^4 \, \big|\overline{S}_{ee} \big|^2 
 \,,\quad 
 T_{2} = \overline{S}_{ee} \overline{S}^*_{\mu \mu} \big|\overline{U}_{\mu 1} \big|^2 \big|\overline{U}_{\mu 2} \big|^2
 \,,\quad 
 T_{3} = \overline{S}_{ee} \overline{S}^*_{\mu e} \big|\overline{U}_{\mu 1} \big|^2 \overline{U}_{\mu 1} \overline{U}^*_{\mu 2} 
\,,\quad 
 T_{4} = \overline{S}_{ee} \overline{S}^*_{e \mu} \big|\overline{U}_{\mu 1} \big|^2 \overline{U}^*_{\mu 1} \overline{U}_{\mu 2} 
    \nonumber \\ 
    && \hspace*{-0.5cm}
 T_{5} = \overline{S}_{\mu \mu } \overline{S}^*_{e e} \big|\overline{U}_{\mu 1} \big|^2 \big|\overline{U}_{\mu 2} \big|^2
 \,,\quad 
 T_{6} =\big|\overline{U}_{\mu 2} \big|^4 \quad \big|\overline{S}_{\mu \mu} \big|^2
 \,,\quad 
 T_{7} =    \overline{S}_{\mu \mu} \overline{S}^*_{\mu e} \big|\overline{U}_{\mu 2} \big|^2 \overline{U}_{\mu 1} \overline{U}^*_{\mu 2}
 \,,\quad 
 T_{8} =
     \overline{S}_{\mu \mu} \overline{S}^*_{e \mu} \big|\overline{U}_{\mu 2} \big|^2 \overline{U}^*_{\mu 1} \overline{U}_{\mu 2}
    \nonumber \\ 
    && \hspace*{-0.5cm} 
  T_{9} =\overline{S}_{\mu e} \overline{S}^*_{e e} \big|\overline{U}_{\mu 1} \big|^2 \overline{U}^*_{\mu 1} \overline{U}_{\mu 2}
 \,,\quad 
  T_{10} =\overline{S}_{\mu e} \overline{S}^*_{\mu \mu} \big|\overline{U}_{\mu 2} \big|^2 \overline{U}^*_{\mu 1} \overline{U}_{\mu 2}
 \,,\quad 
 T_{11} =
    \big|\overline{U}_{\mu 1} \big|^2 \big|\overline{U}_{\mu 2} \big|^2 \big|\overline{S}_{\mu e} \big|^2
 \,,\quad 
 T_{12} =
    \overline{S}_{\mu e} \overline{S}^*_{e \mu} \big(\overline{U}^*_{\mu 1} \big)^2 \big(\overline{U}_{\mu 2} \big)^2
    \nonumber \\ 
&&\hspace*{-0.5cm} T_{13} =
    \overline{S}_{e \mu } \overline{S}^*_{e e} \big|\overline{U}_{\mu 1} \big|^2 \overline{U}^*_{\mu 2} \big(\overline{U}_{\mu 1} \big)^2\,, 
T_{14} =
    \overline{S}_{e \mu } \overline{S}^*_{\mu \mu} \big|\overline{U}_{\mu 2} \big|^2 \overline{U}^*_{\mu 2} \overline{U}_{\mu 1} 
\,, 
T_{15} =
    \overline{S}_{e \mu } \overline{S}^*_{\mu e} \big(\overline{U}_{\mu 1} \big)^2 \big(\overline{U}^*_{\mu 2} \big)^2
\,, T_{16} =
     \big|\overline{U}_{\mu 1} \big|^2 \big|\overline{U}_{\mu 2} \big|^2 \big|\overline{S}_{e \mu} \big|^2 \nonumber \\
     && 
\end{eqnarray} 
\end{widetext}
If we have a closer look at these $16$ terms, either most of them are suppressed by at least by order of $\mathcal{O}(\epsilon^4)$ or combine to form very simple relations. The first term $T_{1} =  
 \big|\overline{U}_{\mu 1} \big|^4 \, \big|\overline{S}_{ee} \big|^2$ is proportional to $\sin^8\theta$ (assuming $\sin \theta_{14}\simeq \sin \theta_{24}$ ) and is suppressed contribution at least by $\mathcal{O}(\epsilon^8)$. We have neglected contributions to the total muon survival probability at the leading order by assuming that either these terms are suppressed by $\mathcal{O}(\epsilon^4)$ or higher order. We can see that terms like $T_1$, $T_2+T_5$, $T_3+T_9$, $T_4+T_{13}$, 
 $T_{11}+T_{16}$ and $T_{12}+T_{15}$ are neglected from the probability analysis. The only non-zero terms contributing to total muon survival probability are given by
\begin{eqnarray}
P^{4\nu}_{\mu \mu}&\approx& 
 \big|\overline{U}_{\mu 2} \big|^4 \quad \big|\overline{S}_{\mu \mu} \big|^2  
 \nonumber \\
 &&+\big|\overline{U}_{\mu 2} \big|^2\, \mbox{Re}\bigg(\overline{S}_{\mu \mu} \,
  \overline{S}^*_{\mu e}\,
  \overline{U}_{\mu 1} \overline{U}^*_{\mu 2}\bigg) \nonumber \\
&&+ \big|\overline{U}_{\mu 2} \big|^2\, \mbox{Re}\bigg(\overline{S}^*_{e \mu} \overline{S}_{\mu \mu} 
  \overline{U}^*_{\mu 1} \overline{U}_{\mu 2}\bigg) \label{muon prob}
\end{eqnarray}  
\noindent
The $T_{6}$, $T_7+T_{10}$, and $T_8+T_{14}$ are the main contributions to the total muon survival probability as first, second and third terms, respectively. The formalism we adopted has effectively simplified the expression for muon survival probability in the $3+1$ flavor mixing scenario under the presence of matter effects.

\begin{table}[ht!]
\begin{center}
\begin{tabular}{lccc}
\hline
\hline
Parameter &  Best Fit values & $1\sigma$   \\
\hline
$\Delta_{21}/10^{-5}~\mathrm{eV}^2 $ (NH ) &7.50  \\
\hline
$\sin^2 \theta_{12}/10^{-1}$ (NH )  & 3.18   \\
\hline
$\Delta_{31}/10^{-3}~\mathrm{eV}^2 $ (NH) & 2.55  \\

\hline
$\sin^2 \theta_{13}/10^{-2}$ (NH) & 2.20  \\

\hline
$\sin^2 \theta_{23}/10^{-1}$ (NH)  & 5.74   \\

\hline
$ \sin^2\theta_{14} $ & 0.02  \\
\hline
$ \sin^2\theta_{24} $  & 0.02  \\
\hline
$ \sin^2\theta_{34} $ &  --   \\
\hline
\hline
\end{tabular}
\caption
{The value of the standard oscillation parameters are taken from the global best fit values quoted in \cite{deSalas:2020pgw}. The value for sterile mixing angles used for carrying out numerical analysis are mentioned alongside.
}
\label{table:1}
\end{center}
\end{table}

\section{Muon survival probability using $\alpha-s13$ approximation}
\noindent
We have the projected Hamiltonian $\overline{H}^{\rm dyn}_{3\nu}$ extracted from complete $4\times4$ Hamiltonian $\overline{H}^{\rm dyn}_{4\nu}$ by considering sterile neutrinos effectively decoupled from the $3+1$ scenario. Also the frequency part of the oscillation probabilities involving $\overline{S}_{ss}$ is averaged out and hence, there is no mass-square difference $\Delta m^2_{41}$. As we are left with an effective $3\times 3$ effective Hamiltonian, we can now easily perform the series expansion. Inspecting the neutrino oscillation parameters taken from the global fit data\cite{deSalas:2020pgw}, one can mark two such oscillations parameters that can be used to carry out the series expansion. These small parameters are as mentioned in the table~\ref{table:1}:
\begin{eqnarray}
\alpha &=& \frac{\Delta m^2_{21}}{\Delta m^2_{31}} \approx 0.026 \nonumber \\
&&\hspace*{-1.0 cm}  s13 = \sin \theta_{13} \approx 0.15
\end{eqnarray}
\noindent
Using the $\alpha-s13$ approximation, the series expansion has already been carried out in the literature~\cite{Akhmedov:2004ny} for the three flavor scheme by expanding them up to second order. Moreover, expansions only in $\alpha$ and $\sin \theta_{13}$ up to the first order have also been performed. But in the presence of sterile neutrinos, it is quite tedious to arrive at the exact analytic formula for muon survival probability keeping terms up to $\alpha^2 \approx {\cal O}(\epsilon^4)$. In this work, we aim to retain terms up to ${\cal O}(\epsilon^{3})$ and neglect all the high order corrections. As we know, $\sin \theta_{13} \approx {\cal O}(\epsilon)$ and sterile mixing angles, $\sin \theta_{i4} \approx {\cal O}(\epsilon)$ (with i=1,2), all these approximations are considered while deriving the final expression for muon survival probability. 
\noindent
The expression for the muon survival probability given in equation ~\ref{muon prob} can actually be simplified further while keeping terms upto order ${\cal O}(\epsilon^{3})$.The resulting terms contributing to the total muon survival probability are expressed as 
\begin{eqnarray}
P^{4\nu}_{\mu \mu} &\approx& P^{3\nu}_{\mu \mu}  + P^{\rm I}_{\rm INT} + P^{\rm II}_{\rm INT} + P^{\rm III}_{\rm INT}
\end{eqnarray}
\noindent
Here the contribution coming from the term $\big|\overline{U}_{\mu 2} \big|^4 \quad \big|\overline{S}_{\mu \mu} \big|^2$ is actually expressed in terms of two factors:
\begin{enumerate}
\item The first one coming from three flavor muon survival probability i.e. $P^{3\nu}_{\mu \mu}$ retaining terms upto ${\cal O}(\epsilon^{3})$.
\item And the other factor $P^{\rm I}_{\rm INT} $ coming from the sterile-active interference term after simplifying the factor $\big|\overline{U}_{\mu 2} \big|^4$ such that terms higher than ${\cal O}(\epsilon^{3})$ are safely neglected.
\end{enumerate}
\noindent
The dominant contribution to muon survival probability independent of any sterile mixing angles and phases is denoted as $P^{3\nu}_{\mu \mu}$. It has been obtained under the vanishing limit of $\alpha ^2$ from the known expression of muon survival probability given in~\cite{Akhmedov:2004ny} and also mentioned in the appendix~\ref{alpha-s13} for completeness.
\begin{eqnarray}
&&\hspace*{-0.5cm} P^{3\nu}_{\mu \mu} \approx 
1 - \sin^22\theta_{23} \sin^2{\Delta} + \alpha \cos^2 \theta_{12} \sin^22\theta_{23} \Delta \sin2\Delta \nonumber \\
              &&\hspace*{-0.5cm} ~ - 4 \sin^2 \theta_{13} \sin^2 \theta_{23} \frac{\sin^2(\hat{A} - 1)\Delta}{(\hat{A} - 1)^2} - \frac{2}{\hat{A} - 1} \sin^2 \theta_{13} \sin^22\theta_{23}  \nonumber \\
              &&\hspace*{-0.5cm} ~ \times \big(\sin\Delta \cos\hat{A}\Delta \frac{\sin(\hat{A} - 1)\Delta}{\hat{A} - 1} - \frac{\hat{A}}{2} \Delta \sin2\Delta \big)  \nonumber \\
              &&\hspace*{-0.5cm}~ - 2 \alpha \sin \theta_{13} \sin2\theta_{12} \sin2\theta_{23} \cos\delta_{CP} \cos\Delta \frac{\sin\hat{A}\Delta}{\hat{A}} \frac{\sin(\hat{A} - 1)\Delta}{\hat{A} - 1}  \nonumber \\
              &&\hspace*{-0.5cm} ~ + \frac{2}{\hat{A} - 1} \alpha \sin \theta_{13} \sin2\theta_{12} \sin2\theta_{23} \cos2\theta_{23} \cos\delta_{CP} \sin\Delta  \nonumber \\
              &&\hspace*{-0.5cm} ~ \times \big(\hat{A} \sin\Delta - \frac{\sin\hat{A}\Delta}{\hat{A}} \cos(\hat{A} - 1) \Delta \big)
              \label{P3nu}
\end{eqnarray}
with $\hat{A} = \frac{A}{\Delta_{31}}$ with $A= 2\sqrt{2} G_F N_e E$ and $\Delta = \Delta_{31} L/4E$, where  $\Delta_{ij} = m_i^2 - m_j^2$, $G_F$ is the Fermi constant, $N_e$ is the electron number density of the medium and $E$ is the energy of the neutrinos.
\newline
\noindent
The first interference term retaining terms $\sin^2\theta_{i4}$ from sterile part while three flavor part is without having any $\alpha$ or $\sin\theta_{13}$ as,
\begin{eqnarray}
&&\hspace*{-0.5cm} P^{\rm I}_{\rm INT} \approx 
  -2 \sin^2\theta_{14}\bigg( 1 - \sin^22\theta_{23} \sin^2{\Delta} \bigg) 
\end{eqnarray}
The other sub-dominant terms are the interference terms $P^{\rm II}_{\rm INT}$ and $P^{\rm III}_{\rm INT}$ corresponding to $\big|\overline{U}_{\mu 2} \big|^2\, \mbox{Re}\big(\overline{S}_{\mu \mu} \, \overline{S}^*_{\mu e}\,
  \overline{U}_{\mu 1} \overline{U}^*_{\mu 2}\big)$ and $\big|\overline{U}_{\mu 2} \big|^2\, \mbox{Re}\big(\overline{S}^*_{e \mu} \overline{S}_{\mu \mu} 
  \overline{U}^*_{\mu 1} \overline{U}_{\mu 2}\big)$ respectively, expanded upto first order in $\sin \theta_{13}$.
\begin{eqnarray}
&&\hspace*{-0.5cm} P^{\rm II}_{\rm INT} \approx 
\frac{1}{4(\hat{A} -1)} \sin \theta_{13} \cos \theta_{24} \sin \theta_{14} \sin \theta_{24} \nonumber \\
&&\hspace*{-0.7 cm}\bigg[\bigg(-3\cos \big(\delta_{14} + (\hat{A} -1)\Delta\big) + \cos \big(\delta_{14} + 2 \hat{A}\Delta +\delta_{13}\big)\bigg)\sin \theta_{23} \nonumber \\
 &&\hspace*{-0.5cm}-2 \sin(\delta_{14}+\delta_{13})\sin \big(2\Delta - \delta_{13}\big) \sin 2 \theta_{23} \nonumber \\
&& \hspace*{-0.5cm}+ 2\cos \big(2\Delta-\delta_{13}\big) \times 
\bigg(\cos \big(\delta_{14} + (2\hat{A} -1)\Delta -\frac{\delta_{13}}{2} \big)\nonumber \\
&&\hspace*{-0.5cm}\sin 3\theta_{23} + \cos \big(\Delta-\frac{\delta_{13}}{2}\big) \cos (\delta_{14}+\delta_{13}) \sin 4\theta_{23} \bigg) \bigg]
\end{eqnarray}

\begin{eqnarray}
&&\hspace*{-1 cm} P^{\rm III}_{\rm INT} \approx 
\frac{1}{2(\hat{A} -1)} \sin \theta_{14} \sin \theta_{13} \sin \theta_{23} \sin \theta_{24} \cos \theta_{24} \nonumber \\
&&\bigg[\cos \bigg(\delta_{14} - 2\Delta\bigg) - \cos(\delta_{14}-\delta_{13}) \nonumber \\
&&+ 2 \cos \big(\Delta -\frac{\delta_{13}}{2}  \big) \cos \big(\delta_{14}-\Delta-\frac{\delta_{13}}{2}  \big) \cos 2\theta_{23}\bigg] \nonumber \\
 && \hspace*{1cm}  
\end{eqnarray}
\noindent
 The important point to notice is that these contributions (i.e.  $P^{\rm II}_{\rm INT}$ and  $P^{\rm III}_{\rm INT}$) actually contain a factor of $\cos^3 \theta_{24} \sin \theta_{14} \sin \theta_{24}$ but since we have consider $\sin \theta_{13}$ and sterile mixing angles, $\sin \theta_{i4}$ (with i=1,2) with order  $\approx {\cal O}(\epsilon)$, therefore we are left with $\cos \theta_{24} \sin \theta_{14} \sin \theta_{24}$.
  
\begin{figure}[hbt]\centering
	\hspace{0.45pt} \includegraphics[width=0.466\textwidth]{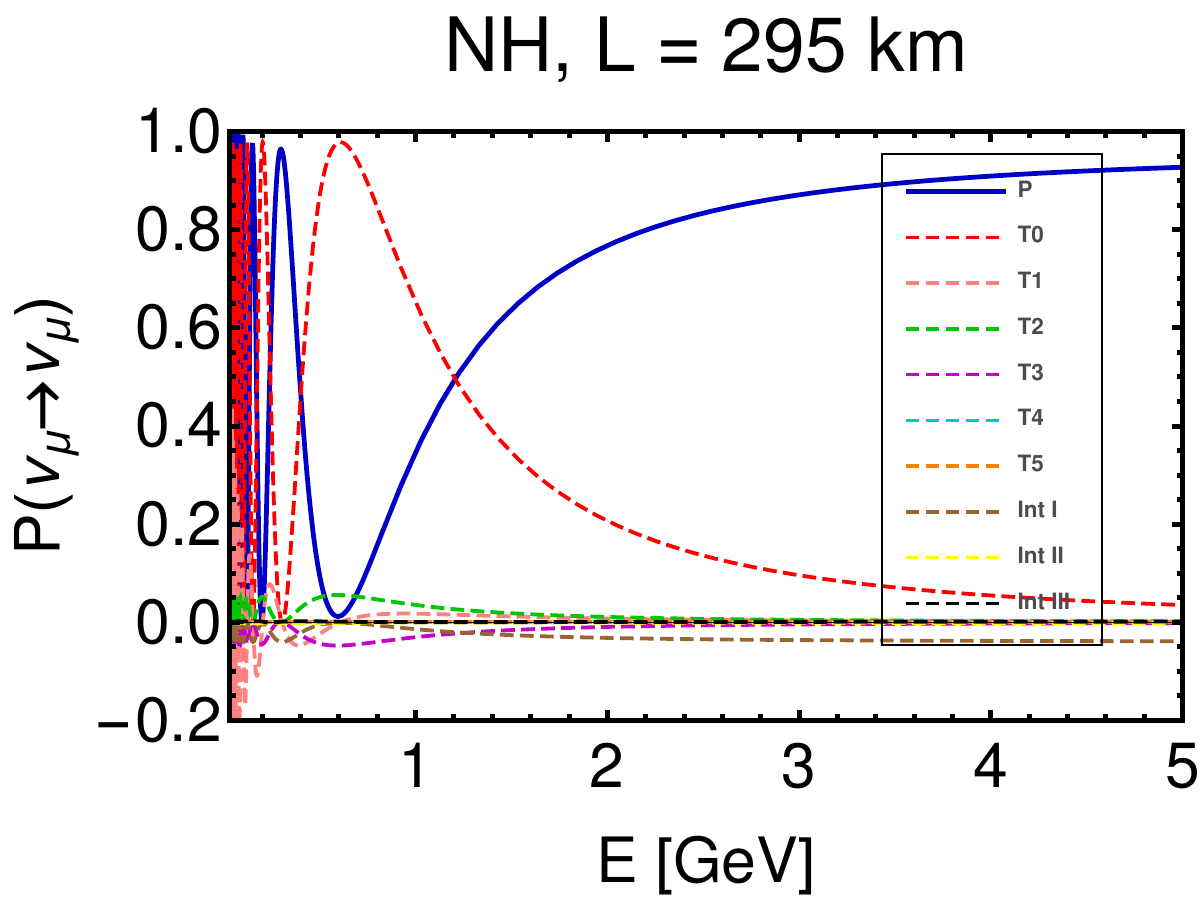}
	\hspace{0.45pt} \includegraphics[width=0.466\textwidth]{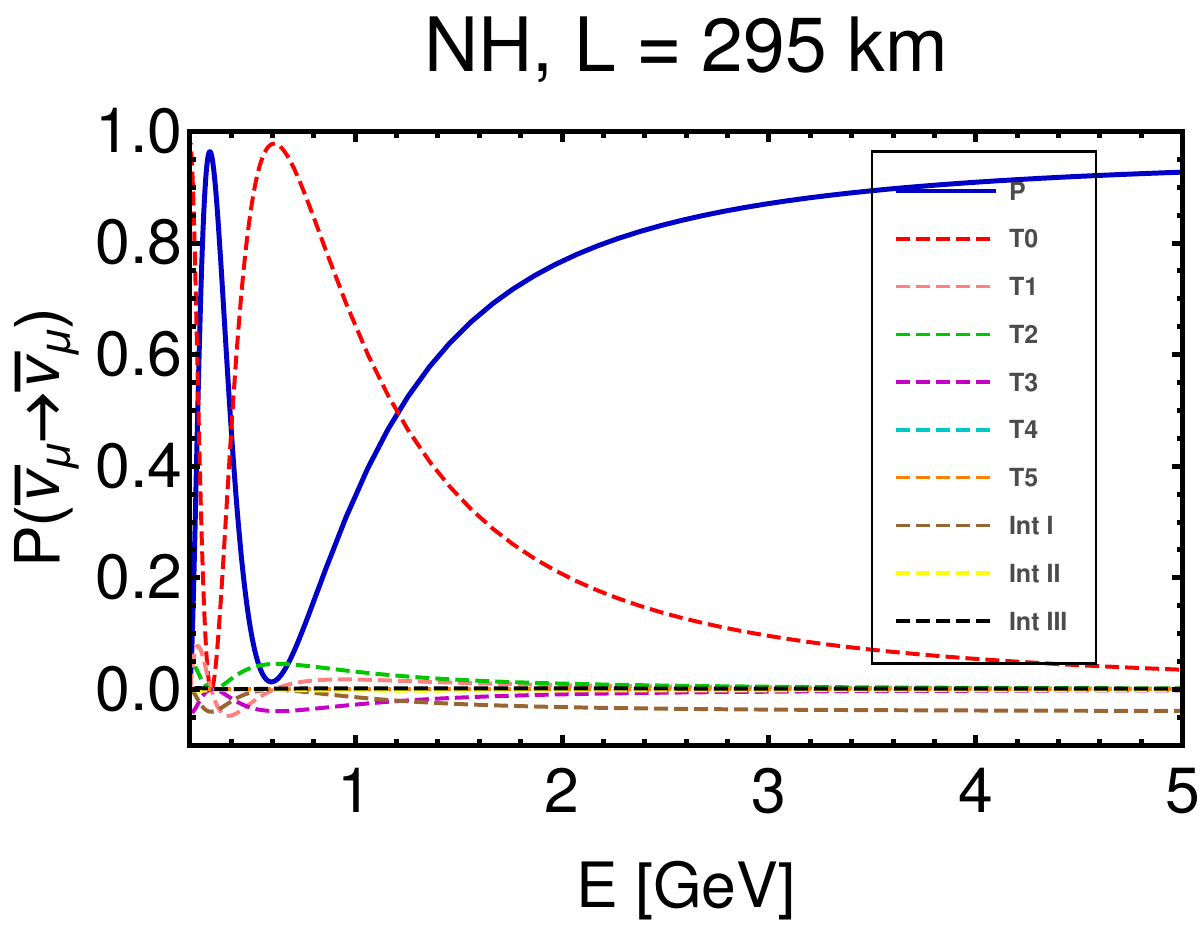}
	\caption{
		The survival probabilities $P_{\nu \mu}$ for neutrino and anti-neutrino mode as a function of neutrino energy in GeV. The baseline is kept fixed at 295 km and the matter density $\rho$ is taken as 2.7g/cc.The sterile mixing angles are assigned value $\sin ^2 \theta_{14} = \sin ^2 \theta_{24} = 0.02$} 
	\label{figure:prob}
\end{figure}  
  
\noindent
To look at the significant contribution coming from the various interference terms (retained upto $\approx {\cal O}(\epsilon ^3)$) involved in the muon survival probability under the $3+1$ scenario, we plot muon survival probability for neutrinos and anti-neutrinos in FIG ~\ref{figure:prob}. The probability variation with the energy of neutrinos is considered under the normal hierarchy for the 295km baseline. The average matter density is taken as $2.7 g/cm^3$ in accordance to PREM profile of Earth matter density~\cite{Dziewonski:1981xy}. The solid blue curve represents the total probability, including all kinds of interferences. The terms marked from $T0-T5$ are the quantified contributions coming from each individual term involved in the equation~\ref{P3nu} for three-flavor probability. The terms Int ${\rm I}$, Int ${\rm II}$ and Int ${\rm III}$ are the subdominant contributions as mentioned in the legend. A similar analysis has been carried out for the anti-neutrinos case by reversing the sign of the fundamental and sterile CP-phases i.e. $\delta_{13}$ and $\delta_{14}$ respectively. Also, the sign of matter potential is changed to look at anti-neutrino probabilities. The value of the mass-squared differences and mixing angles for active neutrinos are kept fixed, as mentioned in Table.1. The active-sterile mixing angles $\sin^2 \theta_{14}$ and $\sin^2 \theta_{24}$ are also kept fixed at $0.02$ throughout the analysis. It appears that the $T0$ term is majorly affecting the total muon probability, with the other terms contributing at sub-leading order. But it is emphasized later how these small contributions arising from active-sterile mixings are widely affecting the behavior under CP and T transformations.
 
\begin{figure}[hbt]\centering
	\hspace{0.45pt} \includegraphics[width=0.466\textwidth]{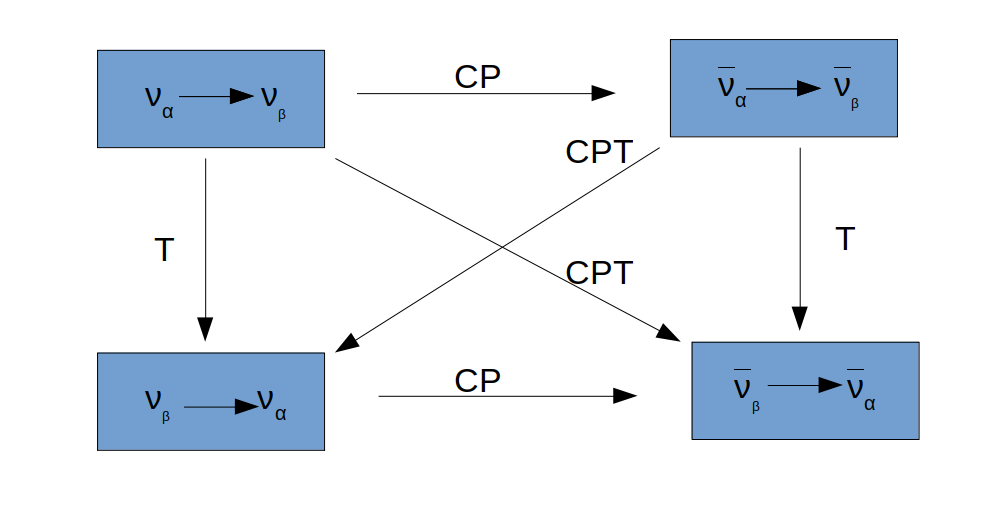}
	\caption{The linkage between CP and T- transformations} 
	\label{figure:CPT}
\end{figure}  
 \noindent 
At the very first look, it may seem that the probability curves are almost identical for both neutrinos and anti- neutrinos. But, zoom in picture marks that they deviate from each other. To pinpoint this difference, which is very important for explaining the matter-antimatter asymmetry, we carry out the CP-violation and T-violation study. The difference between the muon and anti-muon survival probability is defined as $\Delta P_{CP} = P_{\alpha \beta} - P_{\overline{\alpha} \overline{ \beta}}$ and the difference under T transformation is given by $\Delta P_{T} = P_{\alpha \beta} - P_{{\beta} {\alpha}}$. In vacuum, the CPT invariance holds (i.e. $\Delta P_{CP} = \Delta P_{T}$ ) but it is not invariant in presence of matter. But the CPT violation takes place only under an asymmetric density profile. The CP and T transformation linkage has been established in FIG~\ref{figure:CPT}. We are considering constant matter density for our analysis, so any violation of CPT invariance directly hints toward the new physics coming from the sterile presence. 
  \begin{figure}[hbt]\centering
	\hspace{0.45pt} \includegraphics[width=0.466\textwidth]{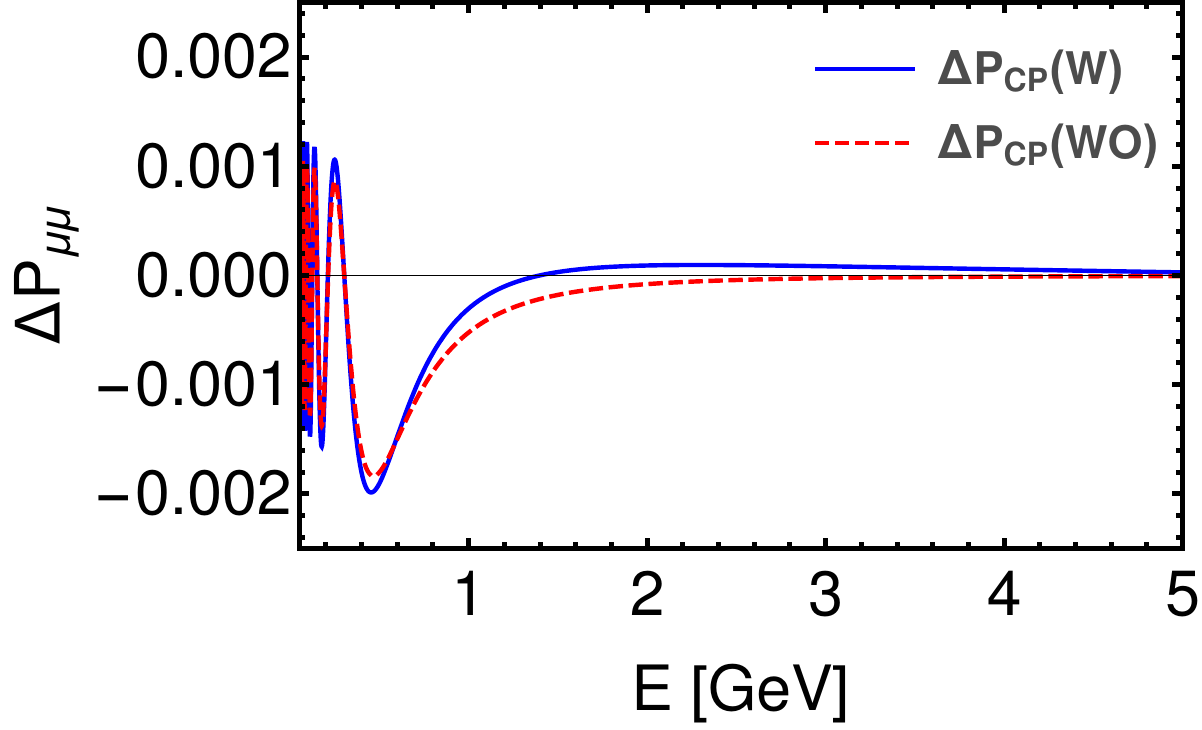}
	\hspace{0.45pt} \includegraphics[width=0.466\textwidth]{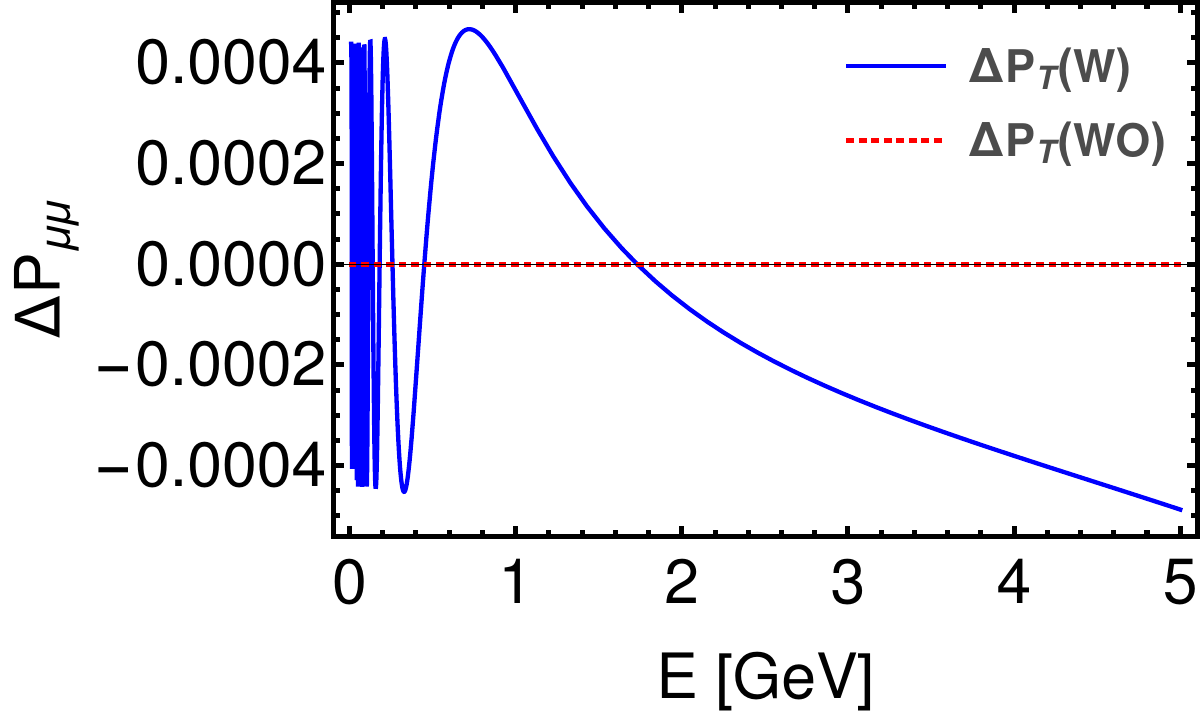}
	\caption{
		The survival probabilities $P_{\nu \mu}$ for neutrino and anti-neutrino mode as a function of neutrino energy in GeV. The baseline is kept fixed at 295 km and the matter density $\rho$ is taken as 2.7g/cc.The sterile mixing angles are assigned value $\sin ^2 \theta_{14} = \sin ^2 \theta_{24} = 0.02$} 
	\label{figure:CPT violation}
\end{figure} 
\newline 

We look at the variation of CP and T violations in FIG ~\ref{figure:CPT violation}. The solid blue curve represents the behavior when all the leading and sub-leading contributions are considered, and it is marked by $\rm {\Delta P_{CP} (W)}$ in legend. The dotted red curve corresponds to the case where we have neglected the contributions from active-sterile interference terms $P^{\rm II}_{\rm INT}$ and $P^{\rm III}_{\rm INT}$ and is represented by $\rm {\Delta P_{CP} (WO)}$. It is interesting to note that the sub-leading terms are playing a significant role towards the CP and T violations,particularly in the case of probability difference under T-transformations. This difference is zero when the presence of active-sterile interference is ignored. Thus, the muon survival probability is highly sensitive to the T-violating effects arising from sterile-neutrino presence. Also, it is evident from the figure that in the presence of matter the CP violation is no more equivalent to the T violation. The violation arises from the two kinds of effects- $\rm i)$ The asymmetry between particle and anti-particle numbers in matter enhances the neutrino oscillations between neutrinos while suppressing between anti-neutrinos. $\rm ii)$ The other one arising from the leptonic CP phase $\delta_{13}$ and $\delta_{14}$. Thus the muon survival probability $P_{\mu \mu}$ plays an important role in understanding the CP-violation and T-violation studies. 
 
 \begin{figure}[hbt]\centering
	\hspace{0.45pt} \includegraphics[width=0.466\textwidth]{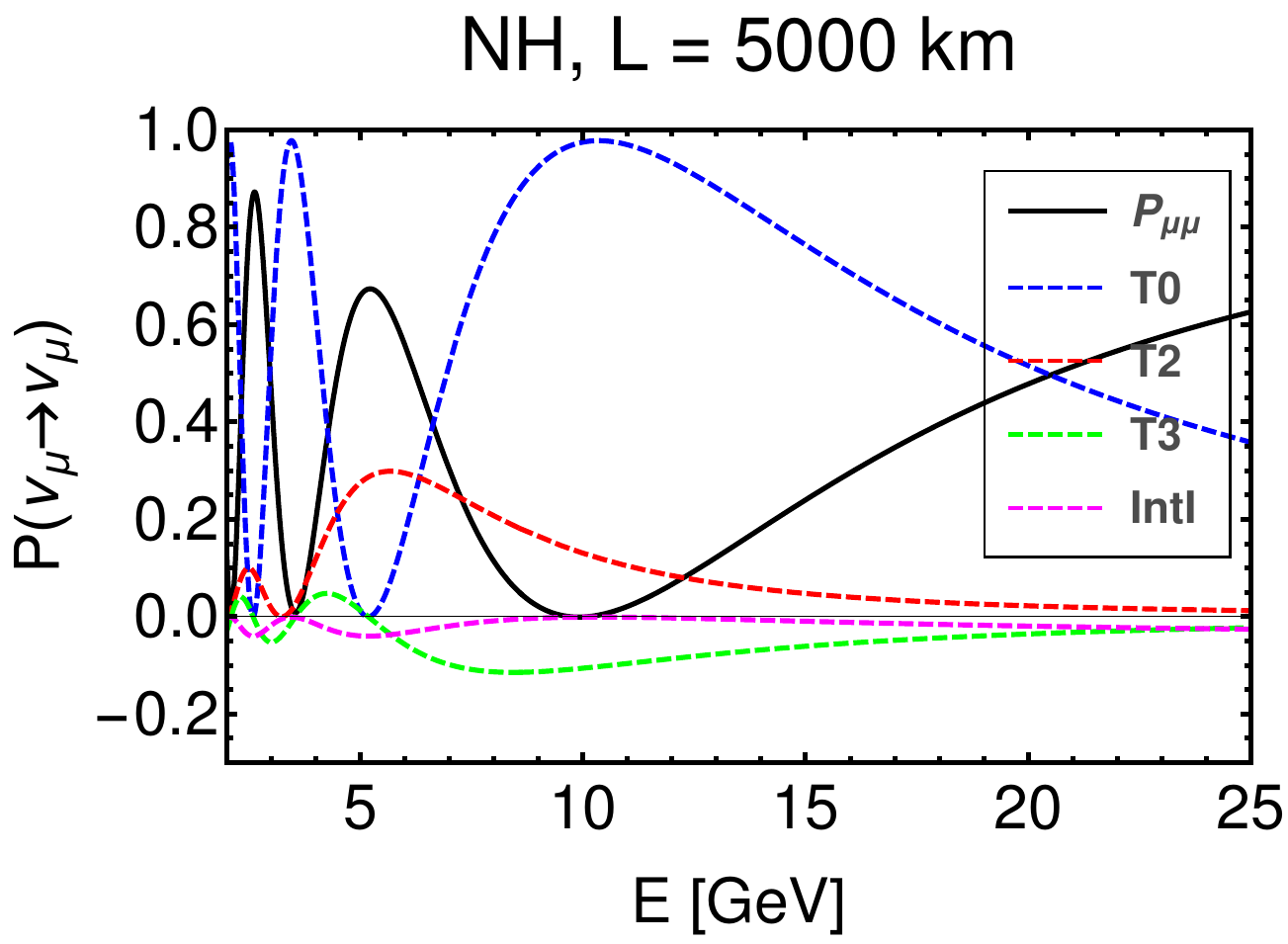}
	\hspace{0.45pt} \includegraphics[width=0.466\textwidth]{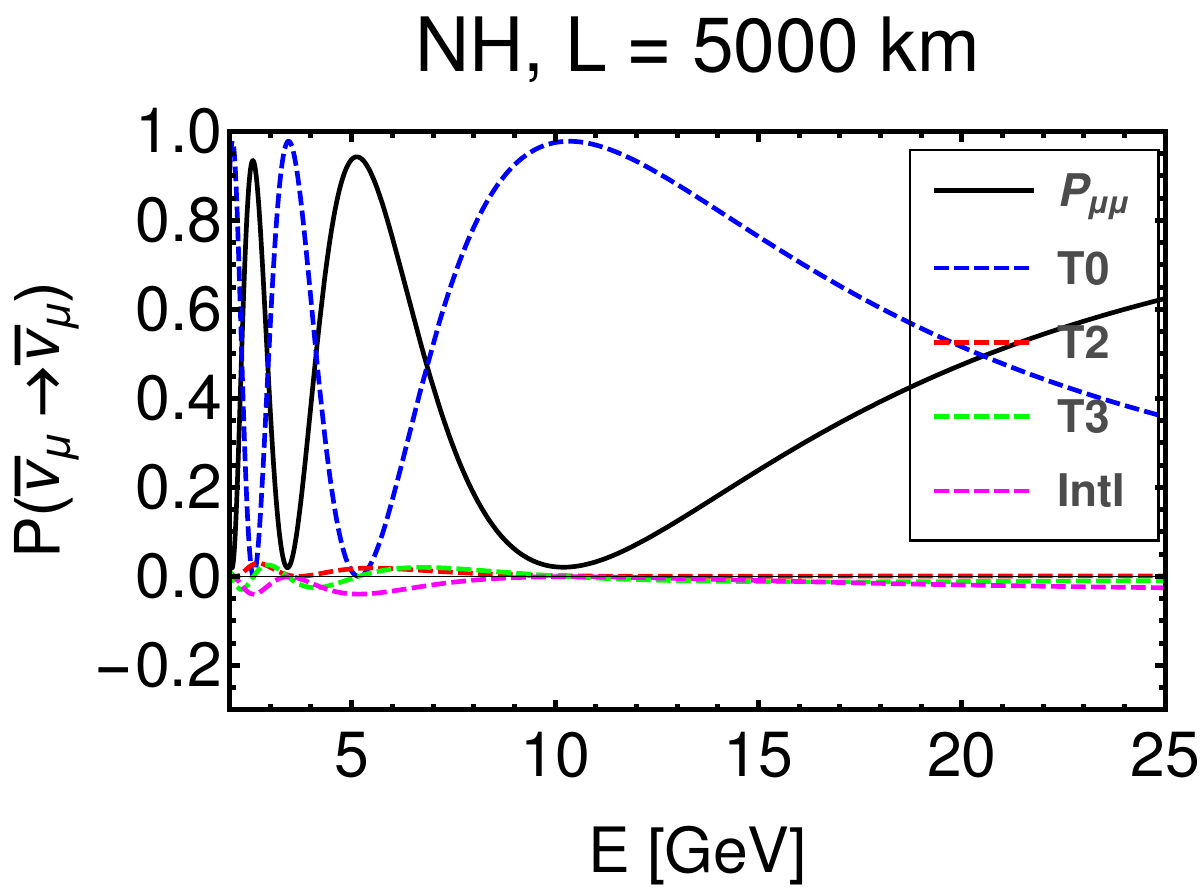}
	\caption{
		The survival probabilities $P_{\nu \mu}$ for neutrino and anti-neutrino mode as a function of neutrino energy in GeV. The baseline is kept fixed at 295 km and the matter density $\rho$ is taken as 2.7g/cc.The sterile mixing angles are assigned value $\sin ^2 \theta_{14} = \sin ^2 \theta_{24} = 0.02$} 
	\label{figure:prob_5000km}
\end{figure}

Furthermore, the muon survival probability acts as an important channel for the long path length/baseline study~\cite{Gandhi:2004bj,Gandhi:2005wa,Gandhi:2007td,Agarwalla:2014fva,Hyper-KamiokandeProto-:2015xww,Agarwalla:2018nlx}. We use our analytic relation for muon survival probability to explore the physics essential at the INO. Since, in the ICAL detector, we have neutrinos with energy in range of 2-20 GeV  over the path lengths (L) varying from 2000 km to 9000 km. At such longer baselines, the effect of the solar mass-squared difference can be safely neglected in comparison with the atmospheric mass-squared difference. Thus, under this approximation the parameter $\alpha$ tends to zero, and we are only left with $T0$, $T2$ and, $T3$ terms in equation~ \ref{P3nu}. Also, neglecting the contribution coming from the active-sterile neutrino interference terms $P^{\rm II}_{\rm INT}$ and $P^{\rm III}_{\rm INT}$, we show the variation of remaining terms in the muon and anti-muon survival probability for baseline $5000$km in FIG~\ref{figure:prob_5000km}. The value of oscillation parameters is as mentioned in table~\ref{table:1}, while the averaged constant matter density is taken as $3.9 g/cc$ based on the PREM profile. The terms $T0$, $T2$, $T3$, and $P^{\rm I}_{\rm INT}$ are represented by dotted blue, red, green, and magenta lines, respectively. The total survival probability is given by solid black lines. The following important conclusions can be drawn from the figure:
\begin{enumerate}
\item The term $T0 = \sin^22\theta_{23} \sin^2{\Delta}$ and $P^{\rm I}_{\rm INT} \approx -2 \sin^2\theta_{14}\bigg( 1 - \sin^22\theta_{23} \sin^2{\Delta} \bigg)$ are independent of any matter contributions, fundamental and sterile CP- phases. As a result, they make an equal contribution to both muon and anti-muon survival probability.
\item The terms $T2 = - 4 s_{13}^2 s_{23}^2 \frac{\sin^2(\hat{A} - 1)\Delta}{(\hat{A} - 1)^2}$, $T3 = - \frac{2}{\hat{A} - 1} s_{13}^2 \sin^22\theta_{23}    \big(\sin\Delta \cos\hat{A}\Delta \frac{\sin(\hat{A} - 1)\Delta}{\hat{A} - 1} - \frac{\hat{A}}{2} \Delta \sin2\Delta \big)$ are dependent on matter term $A$. As the sign of matter contributions are different for neutrino and anti-neutrino interactions, these terms have different contributions in muon and anti-muon survival probability as seen from FIG~\ref{figure:prob_5000km}
\item One also notes that at $E \approx 5 GeV$, the amplitude for total survival probability is more for anti-neutrino mode than neutrino mode. This happens because at this energy value, the terms $T0$ and $T3$ are zero and the survival probability is dependent on $T2$ and $P^{\rm I}_{\rm INT}$ by relation $P_{\mu \mu} = 1 - T2 + P^{\rm I}_{\rm INT}$. Hence, we get a smaller muon survival probability. Similarly, the complete behavior can be explained over the entire energy range by taking the quantitative contributions from each term.
\end{enumerate}
 
 \begin{figure}[hbt]\centering
	\hspace{0.45pt} \includegraphics[width=0.466\textwidth]{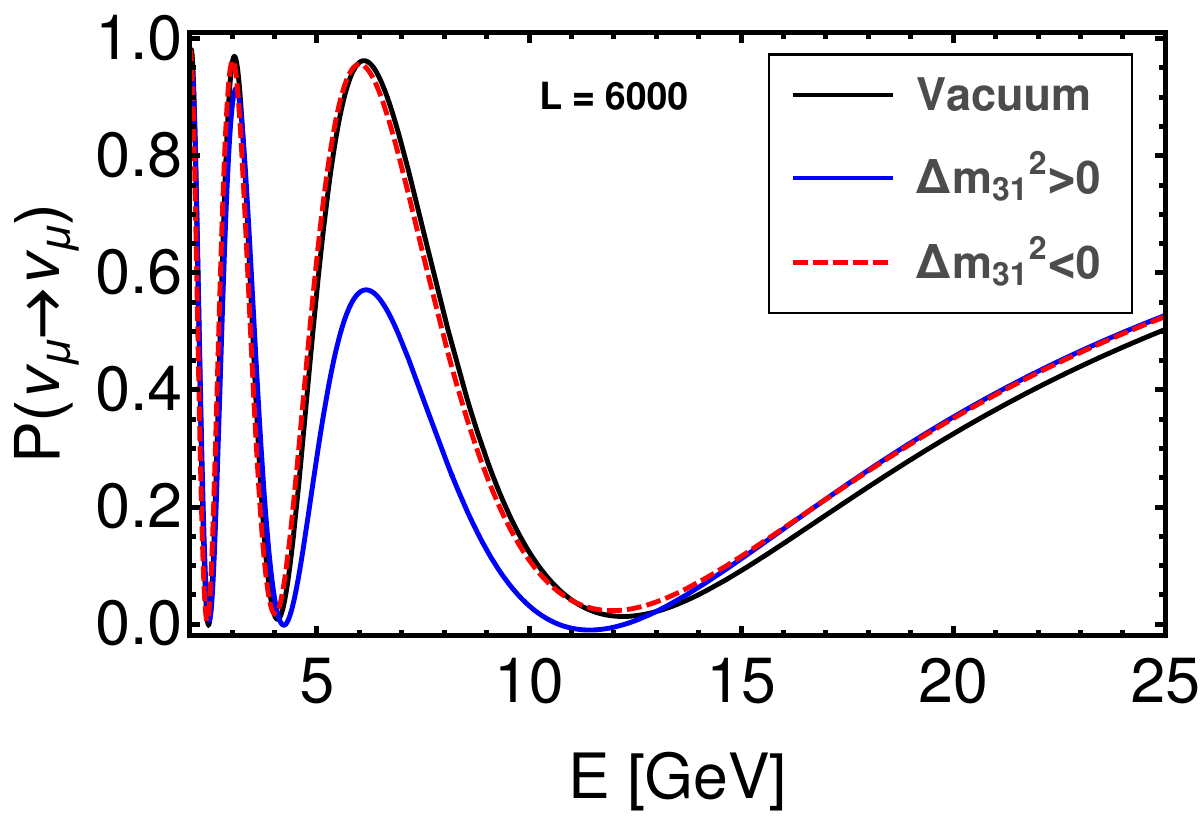}
		\hspace{0.45pt} \includegraphics[width=0.466\textwidth]{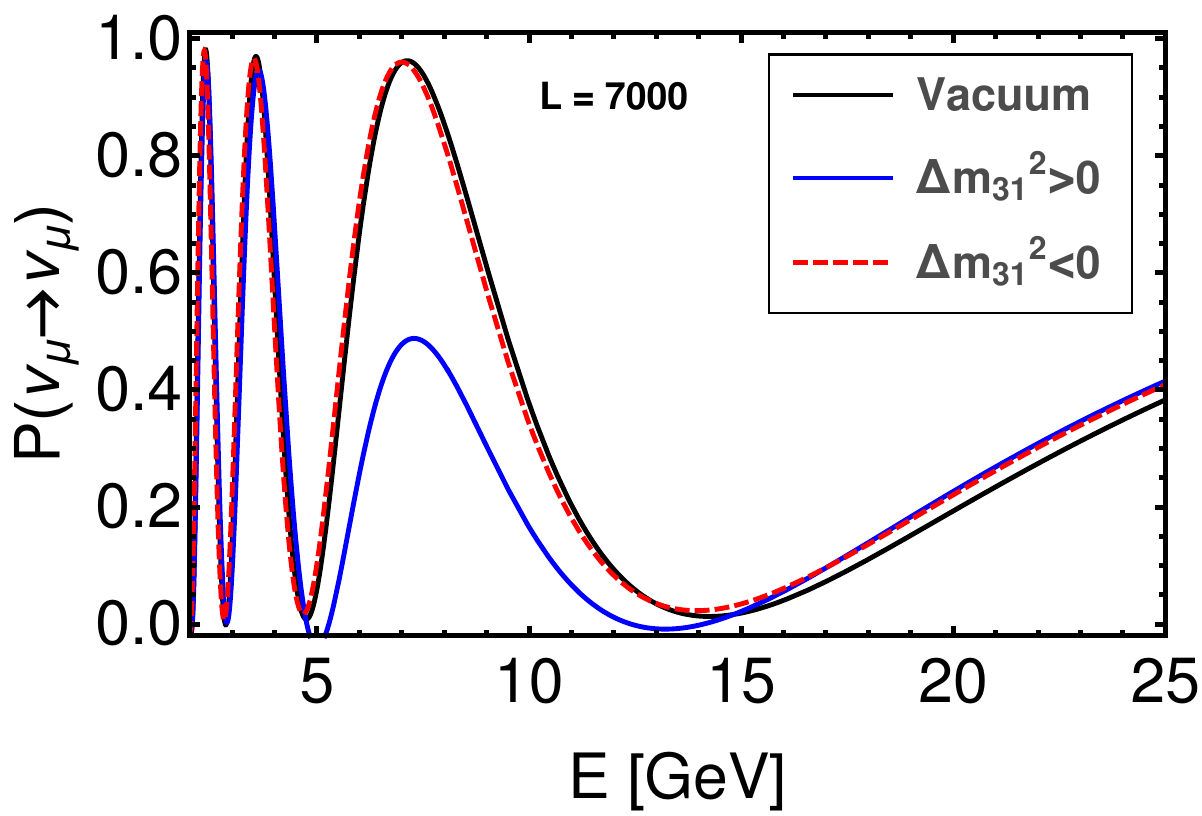}
	\caption{
		The survival probabilities $P_{\nu \mu}$ for neutrino and anti-neutrino mode as a function of neutrino energy in GeV. The baseline is kept fixed at 295 km and the matter density $\rho$ is taken as 2.7g/cc.The sterile mixing angles are assigned value $\sin ^2 \theta_{14} = \sin ^2 \theta_{24} = 0.02$} 
	\label{figure:prob_6000km}
\end{figure}

Let us consider vacuum and matter contributions to muon survival probability relevant for atmospheric neutrino studies in longer baselines at leading order. For such larger baselines, typically of the order of $2000-9000$~km, the contributions involving solar mass-square difference term can be neglected safely and we have ignored the sub-leading sterile neutrino contributions. 
In these limits, the simplified analytic relations for muon survival probability for vacuum and matter effects become,
\begin{eqnarray}
 P^{\rm vac}_{\mu \mu} &=& 1 - \sin^22\theta_{23} \sin^2\Delta + 4 s_{13}^2 s_{23}^2 \cos2\theta_{23} \sin^2\Delta \nonumber \\
 P^{\rm mat}_{\mu \mu} &=& 1 - \sin^22\theta_{23} \sin^2{\Delta}
 - 4 s_{13}^2 s_{23}^2 \frac{\sin^2(\hat{A} - 1)\Delta}{(\hat{A} - 1)^2} \nonumber \\
 &&\hspace*{-0.0 cm}
 - \frac{2}{\hat{A} - 1} \sin_{\theta_{13}}^2 \sin^22\theta_{23}\bigg(\sin\Delta \cos\hat{A}\Delta \frac{\sin(\hat{A} - 1)\Delta}{\hat{A} - 1} 
 \nonumber \\
 &&\hspace*{4cm}- \frac{\hat{A}}{2} \Delta \sin2\Delta \bigg)   
\end{eqnarray}
After a series expansion upto linear order in $\hat{A}$, the matter contribution reduces to,
\begin{eqnarray}
 P^{\rm mat}_{\mu \mu} &=& P^{\rm vac}_{\mu \mu} 
 +8 \hat{A} \cos\big(2\theta_{23}\big) \sin^2\theta_{13} \sin^2\theta_{23} \sin \Delta \nonumber \\
 && \times \big(-\Delta \cos \Delta + \sin \Delta \big)
\end{eqnarray}
\noindent
The atmospheric neutrino probability studies are carried out over the longer baselines where the effect of $\delta_{CP}$ phase is quite insignificant. Taking advantage of this feature of muon survival probability, we look at the variation of probability under normal and inverted mass hierarchy. In FIG ~\ref{figure:prob_6000km}, we show the variation for the baseline of $6000$ km and $7000$ km with the approximate average matter densities $4.1 g/cc$ and $4.15 g/cc$ respectively. A similar analysis can be performed over much longer baselines to build  sharp discrimination between normal and inverted mass hierarchy. However, beyond $10,000$ km, the interferences among the mantle and core arise, and hence such longer baselines are generally avoided for study. The analysis is done for the probability at leading order, independent of active-sterile interferences $P^{\rm II}_{\rm INT}$ and $P^{\rm III}_{\rm INT}$. The solid blue curves represent the behaviour under normal hierarchy, while the dotted red curve represents the inverted hierarchy behaviour. The behaviour in vacuum is marked by a solid black line where we also consider the contribution coming from the $\rm IntI$ term. It is find that the curve for vacuum is overlapping with the probability behaviour under inverted hierarchy, while there is a significant difference between the curves corresponding to both hierarchies. The difference increases with the baseline, therefore it provides a clean way to distinguish between the two hierarchies.

\section{Conclusion}
\noindent
In this paper, we looked at the detailed analytic expressions for the muon survival probability in the $3+1$ scenario under the influence of matter potential. We carry forward the idea of references~\cite{Klop:2014ima, Sharma:2022qeo}, reducing the effective Hamiltonian of the $3+1$ scenario to the projected three- flavor Hamiltonian using S-matrix formalism. We deduced that all the individual contributions to the total muon survival probability could be expressed in just three terms- one leading contribution coming from $3$ flavor muon survival probability and the other two arising from active-sterile mixings. We then performed the well-known $\alpha -s13$ approximation to  further simplify our analytic results. The term by term contributions to the final muon survival probability are quantified numerically for the $T2K$ experiment baseline for both neutrino and anti-neutrino modes. We pointed out that T-violation in muon survival probability arises purely from the new physics of active-sterile mixings. Furthermore, the presence of sterile neutrino highlighted the violation of CPT invariance. We discussed how matter contribution to muon survival probability is related to vaccum contribution at the leading order. 
We also looked at the behaviour of muon probability at the various longer baselines having implications for exploring the atmospheric neutrino sector. Moreover, we discuss the role of muon survival probability in resolving the mass-hierarchy degeneracy. 
\section{Acknowledgment}
Kiran Sharma would like to acknowledge the Ministry of Education, Govt of India for financial support.

\section{APPENDIX}
 
\appendix

\section{Parametrization of mixing matrix in 3+1 framework.}
\label{app:U44}
The standard parametrization considered for the present work is given by 
\begin{eqnarray}
U &=& R\big(\theta_{34}, \delta_{34} \big) \, R\big(\theta_{24}, 0 \big) \, R\big(\theta_{14}, \delta_{14} \big) \nonumber \\
 &&\hspace*{1cm} \times R\big(\theta_{23}, 0 \big) R\big(\theta_{13}, \delta_{13}\big) R\big(\theta_{12}, 0 \big) 
\end{eqnarray}
Here, $\theta_{14}$, $\theta_{24}$ and $\theta_{34}$ are the mixing angles between sterile and active neutrinos while $\theta_{23}$, $\theta_{13}$ and $\theta_{12}$ are known mixing angles of three flavor neutrinos. The fundamental CP phase $\delta_{13}$ and other CP-phases $\delta_{14}$ and $\delta_{34}$ coming from active-sterile mixings. One can also write down the complete $4\times4$ mixing matrix in terms of usual three flavor matrix as,
\begin{eqnarray}
\hspace*{-1.5cm} U = R\big(\theta_{34}, \delta_{34} \big) \, R\big(\theta_{24}, 0\big) \, R\big(\theta_{14}, \delta_{14} \big) U_{3\nu} 
\end{eqnarray}
$\quad$ with $U_{3\nu}= R\big(\theta_{23}, 0 \big) R\big(\theta_{13}, \delta_{13}\big) R\big(\theta_{12}, 0 \big)$.

The general matrix structure of $U$ and $U^\dagger$ can be read as,
\begin{eqnarray}
 U &=& \begin{pmatrix}
        U_{e1} & U_{e2} & U_{e3} & U_{e4} \\
        U_{\mu 1} & U_{\mu 2} & U_{\mu 3} & U_{\mu 4} \\
        U_{\tau 1} & U_{\tau 2} & U_{\tau 3} & U_{\tau 4} \\
        U_{s1} & U_{s2} & U_{s3} & U_{s4} 
       \end{pmatrix} \, ,
  U^\dagger  = \begin{pmatrix}
        U^*_{e1} & U^*_{\mu 1} & U^*_{\tau 1} & U^*_{s1} \\
        U^*_{e2} & U^*_{\mu 2} & U^*_{\tau 2} & U^*_{s2} \\
        U^*_{e3} & U^*_{\mu 3} & U^*_{\tau 3} & U^*_{s3} \\
        U^*_{e4} & U^*_{\mu 4} & U^*_{\tau 4} & U^*_{s4} 
       \end{pmatrix} \nonumber 
\end{eqnarray}

The relation $UU^\dagger$ gives,
\begin{eqnarray}
&& \hspace*{-0.5cm} \left[\begin{array}{cccc}
\sum_{j=1}^4 U_{ei} U^*_{ei} & \sum_{j=1}^4 U_{ei} U^*_{\mu i} & \sum_{j=1}^4 U_{ei} U^*_{\tau i} & \sum_{j=1}^4 U_{ei} U^*_{si} \\
\sum_{j=1}^4 U_{\mu i} U^*_{e i} & \sum_{j=1}^4 U_{\mu i} U^*_{\mu i} & \sum_{j=1}^4 U_{\mu i} U^*_{\tau i} & \sum_{j=1}^4 U_{\mu i} U^*_{si} \\
\sum_{j=1}^4 U_{\tau i} U^*_{ei} & \sum_{j=1}^4 U_{\tau i} U^*_{\mu i} & \sum_{j=1}^4 U_{\tau i} U^*_{\tau i} & \sum_{j=1}^4 U_{\tau i} U^*_{si} \\
\sum_{j=1}^4 U_{si} U^*_{ei} & \sum_{j=1}^4 U_{si} U^*_{\mu i} & \sum_{j=1}^4 U_{si} U^*_{\tau i} & \sum_{j=1}^4 U_{si} U^*_{si} \\ 
\end{array}\right] \nonumber \\
&&\hspace*{2cm} = \left[ \begin{array}{cccc} 1 & 0 & 0 & 0 \\
                          0 & 1 & 0 & 0 \\
                          0 & 0 & 1 & 0 \\
                          0 & 0 & 0 & 1
       \end{array}
\right]
\label{Urel}
\end{eqnarray}
The unitarity condition of the PMNS mixing matrix in $3+1$ framework leads to
 $$\sum^{4}_{j=1} U_{\alpha j} U^*_{\beta j} = \delta_{\alpha \beta}= \bigg\{\begin{array}{cc}
                                                    1 \quad \mbox{if $\alpha = \beta$}\\ 0 \quad \mbox{if $\alpha \neq \beta$}
                                                   \end{array}
  $$
The rowwise elements of the complete $4\times4$ mixing matrix are expressed as follows:
\begin{enumerate}
 \item[I.] {\bf First  column elements of $U$:} 
 \begin{itemize}
  \item[] 
  \begin{eqnarray}
   &&U_{e1}= c_{21} c_{13} c_{14}  \nonumber \\
   &&U_{\mu 1} = c_{12} \left(-c_{13} s_{14} s_{24} e^{\text{i$\delta $}_{14}}-c_{24} s_{13} s_{23} e^{\text{i$\delta $}_{13}}\right)-c_{23} c_{24} s_{12} \nonumber           \\
   &&U_{\tau 1} = c_{12} \bigg(-c_{13} c_{24} s_{14} s_{34} e^{\text{i$\delta $}_{14}-\text{i$\delta $}_{34}}  \nonumber \\ && \hspace*{1.5cm}
   -s_{13} e^{\text{i$\delta $}_{13}} \left(c_{23} c_{34}-s_{23} s_{24} s_{34} e^{-\text{i$\delta $}_{34}}\right)\bigg) \nonumber \\ && \hspace*{1.5cm}
   -s_{12} \left(-c_{23} s_{24} s_{34} e^{-\text{i$\delta $}_{34}}-c_{34} s_{23}\right)
   \nonumber \\ &&U_{s 1} = 
   c_{12} \bigg(-s_{13} e^{\text{i$\delta $}_{13}} \left(-c_{23} s_{34} e^{\text{i$\delta $}_{34}}-c_{34} s_{23} s_{24}\right)  \nonumber \\ && \hspace*{-0.1cm}
   -c_{13} c_{24} c_{34} s_{14} e^{\text{i$\delta $}_{14}}\bigg) 
   -s_{12} \left(s_{23} s_{34} e^{\text{i$\delta $}_{34}}-c_{23} c_{34} s_{24}\right) 
  \end{eqnarray}
 \end{itemize}
 \item[II.] {\bf Second  column elements of $U$:} 
 \begin{itemize}
  \item[] 
  \begin{eqnarray}
   &&U_{e2}= c_{13} c_{14} s_{12}  \nonumber \\
   &&U_{\mu 2} = s_{12} \left(-c_{13} s_{14} s_{24} e^{\text{i$\delta $}_{14}}-c_{24} s_{13} s_{23} e^{\text{i$\delta $}_{13}}\right)+c_{12} c_{23} c_{24} \nonumber           \\
   &&U_{\tau 2} = c_{12} \left(-c_{23} s_{24} s_{34} e^{-\text{i$\delta $}_{34}}-c_{34} s_{23}\right) \nonumber \\ && \hspace*{1.5cm}
   +s_{12} \bigg(-c_{13} c_{24} s_{14} s_{34} e^{\text{i$\delta $}_{14}-\text{i$\delta $}_{34}} \nonumber \\ && \hspace*{1.5cm}
   -s_{13} e^{\text{i$\delta $}_{13}} \left(c_{23} c_{34}-s_{23} s_{24} s_{34} e^{-\text{i$\delta $}_{34}}\right)\bigg)
   \nonumber \\ &&U_{s 2} = c_{12} \left(s_{23} s_{34} e^{\text{i$\delta $}_{34}}-c_{23} c_{34} s_{24}\right) \nonumber \\ && \hspace*{1.5cm}
   +s_{12} \bigg(-s_{13} e^{\text{i$\delta $}_{13}} \left(-c_{23} s_{34} e^{\text{i$\delta $}_{34}}-c_{34} s_{23} s_{24}\right) \nonumber \\ && \hspace*{1.5cm}
   -c_{13} c_{24} c_{34} s_{14} e^{\text{i$\delta $}_{14}}\bigg)
  \end{eqnarray}
 \end{itemize}
  \item[III.] {\bf Third  column elements of $U$:} 
 \begin{itemize}
  \item[] 
  \begin{eqnarray}
   &&U_{e3}= c_{14} s_{13} e^{-\text{i$\delta $}_{13}} \nonumber \\
   &&U_{\mu 3} = c_{13} c_{24} s_{23}-s_{13} s_{14} s_{24} e^{\text{i$\delta $}_{14}-\text{i$\delta $}_{13}} \nonumber           \\
   &&U_{\tau 3} =c_{13} \left(c_{23} c_{34}-s_{23} s_{24} s_{34} e^{-\text{i$\delta $}_{34}}\right) \nonumber \\ && \hspace*{1.5cm}
   -c_{24} s_{13} s_{14} s_{34} e^{-\text{i$\delta $}_{13}+\text{i$\delta $}_{14}-\text{i$\delta $}_{34}} \nonumber           \\
   &&U_{s 3} = c_{13} \left(-c_{23} s_{34} e^{\text{i$\delta $}_{34}}-c_{34} s_{23} s_{24}\right) \nonumber \\ && \hspace*{1.5cm}
   -c_{24} c_{34} s_{13} s_{14} e^{\text{i$\delta $}_{14}-\text{i$\delta $}_{13}}     
 \end{eqnarray}
 \end{itemize}
  \item[IV.] {\bf Fourth column elements of $U$:} 
 \begin{itemize}
  \item[] 
  \begin{eqnarray}
   &&U_{e4}= s_{14} e^{-\text{i$\delta $}_{14}}  \nonumber \\
   &&U_{\mu 4} = c_{14} s_{24} \nonumber           \\
   &&U_{\tau 4} =c_{14} c_{24} s_{34} e^{-\text{i$\delta $}_{34}} \nonumber           \\
   &&U_{s 4} =c_{14} c_{24} c_{34}          
  \end{eqnarray}
 \end{itemize}
\end{enumerate}

\section{Three flavor oscillation probabilities in presence of matter using $\alpha-s13$ approximations:}
\label{alpha-s13}
We will adopt the formalism of $\alpha-\sin\theta_{13}$ approximation as discussed in ref ~\cite{Akhmedov:2004ny} to derive the muon survival probability for three flavor neutrino oscillation. Usually, the oscillation probability is expressed as a series expansion up to $\alpha^2$ with $\alpha=\Delta m^2_{21}/\Delta m^2_{31}$. Let us define three key parameters which we will use in expressing muon survival probability in presence of matter,
\begin{tcolorbox}[colback=white!10,colframe=black!40!black,title=]
\begin{itemize}
 \item $\Delta=\frac{\Delta m^2_{31}L}{4E}$.
 \item $\alpha = \Delta m^2_{21} / \Delta m^2_{31}$.
 \item $\left(\alpha \Delta \right) = \Delta m^2_{21}$.
\end{itemize}
\end{tcolorbox}
The effective Hamiltonian in flavor basis can be written as
 \begin{eqnarray}
  H = \frac{\Delta_{31}}{2E} \big[ U \text{diag}(0, \alpha, 1) U^\dagger + \text{diag}(\hat{A}, 0, 0) \big],  
 \end{eqnarray}
 where $\hat{A} = A/\Delta_{31}$. In order to derive the double expansion, we write the above Hamiltonian as
 \begin{eqnarray}
  H = \frac{\Delta_{31}}{2E} R_{23} U_{\delta} M U_{\delta}^\dagger R_{23}^T,
 \end{eqnarray}
 where $U_{\delta} = \text{diag}(1, 1, e^{i \delta_{CP}})$. We define, 
\begin{eqnarray}
H &=& \frac{\Delta_{31}}{2E} M \nonumber \\
  &=& \frac{\Delta_{31}}{2E} \left[R_{13} R_{12} \text{diag}(0, \alpha, 1) R_{12}^T R_{13}^T + \text{diag} (\hat{A}, 0, 0)\right] \nonumber \\
    &=&
    \begin{pmatrix}
     s_{12}^2 c_{13}^2 \alpha + s_{13}^2 + \hat{A} & \alpha c_{12} c_{13} s_{12} & s_{13} c_{13} (1 - \alpha s_{12}^2) \\
    s_{12} c_{12} c_{13} \alpha & \alpha c_{12}^2 & -\alpha c_{12} s_{12} s_{13} \\
    s_{13} c_{13} (1 - \alpha s_{12}^2) & - s_{12} c_{12} s_{13} \alpha & \alpha s_{12}^2 s_{13}^2 + c_{13}^2 \\
    \end{pmatrix}.
\end{eqnarray}
To start with the diagonalization of the above mass matrix, let us make the desired approximation that $\alpha \approx \sin\theta_{13} \approx \epsilon$ for a small parameter $\epsilon$. Although this approximation is not entirely correct as $\alpha \approx \pm 0.03$ while $\sin\theta_{13} \equiv s_{13} \approx 0.15$. Also there are other ways to find oscillation probabilities using expansion that accounts for different order of $\alpha$ and $\sin\theta_{13}$ ~\cite{Asano:2011nj,Agarwalla:2013tza}. However, these parameters are being small and similar in order of magnitude, the simple second order expansion in terms of $\epsilon$ is still valid. Using perturbation theory up to second order in the small parameters $\alpha$ and $s_{13}$, the resulting energy eigenvalues ( $E_i = \frac{\Delta_{31}}{2E} \lambda_i$) are given 
by
\begin{eqnarray} \label{energy_1}
 E_1  &=& \frac{\Delta_{31}}{2E} \big(\hat{A} + \alpha s_{12}^2 + s_{13}^2 \frac{\hat{A}}{\hat{A}-1} + \alpha^2 \frac{\sin^22\theta_{12}}{4\hat{A}} \big), 
 \nonumber \\ \label{energy_2}
 E_2 &=& \frac{\Delta_{31}}{2E} \big( \alpha c_{12}^2 - \alpha^2 \frac{\sin^22\theta_{12}}{4\hat{A}} \big), \nonumber \\ \label{energy_3}
 E_3 &=& \frac{\Delta_{31}}{2E} \big( 1 - s_{13}^2 \frac{\hat{A}}{\hat{A} - 1} \big),
\end{eqnarray}
Similarly, the resulting eigenvectors derived using the same approximation are as follows,
 \begin{eqnarray}
 &&v_1 = 
 \begin{pmatrix}
  1 \\
  \frac{\alpha \sin2\theta_{12}}{2\hat{A}} +\frac{\alpha^2 \sin4\theta_{12}}{4\hat{A}^2} \\
  \frac{s_{13}}{\hat{A} - 1} - \frac{\hat{A} \alpha s_{13} s_{12}^2}{(\hat{A} -1)^2} \\
 \end{pmatrix}, \hspace{3 mm} \nonumber \\
&& v_2 = 
 \begin{pmatrix}
  -\frac{\alpha \sin2\theta_{12}}{2\hat{A}} - \frac{\alpha^2 \sin4\theta_{12}}{4 \hat{A}^2} \\
  1 \\
  \frac{\alpha s_{13} \sin2\theta_{12} (\hat{A}+1)}{2\hat{A}} \\
 \end{pmatrix}, \nonumber  \\ 
 &&v_3 =
 \begin{pmatrix}
  -\frac{s_{13}}{\hat{A} - 1} + \frac{\hat{A} \alpha s_{13} s_{12}^2}{(\hat{A} - 1)^2} \\
  \frac{\hat{A} \alpha s_{13} \sin2\theta_{12}}{2(\hat{A} - 1)} \\
  1 \\
 \end{pmatrix}.
\end{eqnarray}
By stacking these eigenvectors, one can obtain the relevant mixing matrix is $W = (v_1, v_2, v_3)$ and the modified mixing matrix is read as,
\begin{eqnarray}
 U_M = R_{23} U_{\delta} W,
\end{eqnarray}
\noindent
Let us now derive oscillation probability using calculated energy eigenvalues and modified mixing matrix. With some simplifications, the muon survival probability is found to be,
\begin{eqnarray}
 P_{\mu \mu} &=& 1 - \sin^22\theta_{23} \sin^2{\Delta} + \alpha c_{12}^2 \sin^22\theta_{23} \Delta \sin2\Delta \nonumber \\ 
              &&\hspace*{-0.5 cm} ~ - \alpha^2 \sin^22\theta_{12}c_{23}^2 \frac{\sin^2\hat{A} \Delta}{\hat{A}^2} - \alpha^2 c_{12}^2 \sin^22\theta_{23} \Delta^2 \cos2\Delta \nonumber \\  
              &&\hspace*{-0.5 cm} ~ + \frac{1}{2\hat{A}} \alpha^2 \sin^22\theta_{12} \sin^22\theta_{23} \nonumber \\ 
              &&\hspace*{-0.5 cm} ~ \times \big(\sin\Delta \frac{\sin\hat{A}\Delta}{\hat{A}} \cos(\hat{A} - 1)\Delta - \frac{\Delta}{2} \sin2\Delta \big) \nonumber \\ 
              &&\hspace*{-0.5 cm} ~ - 4 s_{13}^2 s_{23}^2 \frac{\sin^2(\hat{A} - 1)\Delta}{(\hat{A} - 1)^2} - \frac{2}{\hat{A} - 1} s_{13}^2 \sin^22\theta_{23} \nonumber \\ 
              &&\hspace*{-0.5 cm} ~ \times \big(\sin\Delta \cos\hat{A}\Delta \frac{\sin(\hat{A} - 1)\Delta}{\hat{A} - 1} - \frac{\hat{A}}{2} \Delta \sin2\Delta \big) \nonumber \\ 
              &&\hspace*{-1 cm} ~ - 2 \alpha s_{13} \sin2\theta_{12} \sin2\theta_{23} \cos\delta_{CP} \cos\Delta \frac{\sin\hat{A}\Delta}{\hat{A}} \frac{\sin(\hat{A} - 1)\Delta}{\hat{A} - 1} \nonumber \\ 
              && \hspace*{-0.5 cm}~ + \frac{2}{\hat{A} - 1} \alpha s_{13} \sin2\theta_{12} \sin2\theta_{23} \cos2\theta_{23} \cos\delta_{CP} \sin\Delta \nonumber \\ 
              && \hspace*{-0.5 cm}~ \times \big(\hat{A} \sin\Delta - \frac{\sin\hat{A}\Delta}{\hat{A}} \cos(\hat{A} - 1) \Delta \big),
\end{eqnarray}
with $\alpha = \Delta_{21}/\Delta_{31}$, $\hat{A} = A/\Delta_{31}$, $A= 2\sqrt{2} G_F N_e E$ and $\Delta = \Delta_{31}/4E$.
Where  $\Delta_{ij} = m_i^2 - m_j^2$, $G_F$ is the Fermi constant, $N_e$ is the electron number density of the medium and $E$ is the energy of the neutrinos.
\newline
\noindent
The vacuum oscillation probabilities up to second order in $\alpha$ and $s_{13}$ can be readily derived with the approximation $\hat{A} \rightarrow 0$ in the 
above set of equation.
\begin{eqnarray}
 P_{\mu \mu} &=& 1 - \sin^22\theta_{23} \sin^2\Delta + \alpha c_{12}^2 \sin^2\theta_{23} \Delta \sin2\Delta \nonumber \\ 
             && ~ - \alpha^2 \Delta^2 \big[\sin^22\theta_{12} c_{23}^2 + c_{12}^2 \sin^22\theta_{23} \big(\cos2\Delta - s_{12}^2 \big) \big] \nonumber \\ 
             && ~ + 4 s_{13}^2 s_{23}^2 \cos2\theta_{23} \sin^2\Delta \nonumber \\ 
             && ~ - 2 \alpha s_{13} \sin2\theta_{12} s_{23}^2 \sin2\theta_{23} \cos\delta_{CP} \Delta \sin2\Delta,
\end{eqnarray}
 
\bibliographystyle{apsrev4-1}
\bibliography{neutrino.bib}
\end{document}